# Implementing dynamic high-performance computing supported workflows on Scanning Transmission Electron Microscope

Utkarsh Pratiush[1], Austin Houston[1], Sergei V Kalinin[1,2], and Gerd Duscher[1]

[1] Department of Materials Science and Engineering, University of Tennessee, Knoxville, TN 37996, USA

[2] Pacific Northwest National Laboratory, Richland, WA 99354

**Abstract**

Scanning Transmission Electron Microscopy (STEM) coupled with Electron Energy Loss Spectroscopy (EELS) presents a powerful platform for detailed material characterization via rich imaging and spectroscopic data. Modern electron microscopes can access multiple length scales and sampling rates far beyond human perception and reaction time. Recent advancements in machine learning (ML) offer a promising avenue to enhance these capabilities by integrating ML algorithms into the STEM-EELS framework, fostering an environment of active learning. This work enables the seamless integration of STEM with High-Performance Computing (HPC) systems. We present several implemented workflows that exemplify this integration. These workflows include sophisticated techniques such as object finding and Deep Kernel Learning (DKL). Through these developments, we demonstrate how the fusion of STEM-EELS with ML and HPC enhances the efficiency and scope of material characterization for 70% STEM available globally. The codes are available at [GitHub link](GitHub link).

I. Introduction

Scanning Electron Transmission Microscopy (STEM) has become one of foundational tools in materials science, condensed matter physics, chemistry, catalysis and other fields. STEM allows imaging and characterization of materials at nanoscale extending to atomic resolution, providing insights into the atomic and molecular structures of materials. By now, STEM has become one of the primary tools in multiple academic and industrial research labs worldwide[1–6].

The versatility of STEM is further enhanced by its integration with techniques such as electron energy loss spectroscopy (EELS)[78–11], technique which allows for precise analysis of the chemical composition[12], electronic structure[13] of materials, and low energy quasiparticles[14]. The ability of STEM to image materials at nanometer and atomic levels makes it a crucial tool for advancing our understanding of the structure property relationship in wide range of material systems. This combination is particularly beneficial in the development of new materials for advanced technological applications, including semiconductors, solarcells[15], catalysts[16], and battery materials[17].

Like other domains, the growth of data science and machine learning have stimulated the interest to the big data and machine learning methods in the electron microscopy. A number of early works have been reported in 80ies and 90ies, including that of Duscher[18] and Bonnett[19]. However, at that time, the computational capabilities and hence potential to work with large data volumes have been limited, as were the library of available data tools.

The first sustained effort in ML for EM can be dated to the work of Watanabe et al[20] in ~2005. At that time, the grid-based spectroscopic measurements had become common on many tools, and the computation power and codes sufficient for simple data analytics have become available to broad electron microscopy community. From that $_{moment}$, the combination of the hyperspectral imaging and machine learning for physics extraction and dimensionality reduction has become the new paradigm in STEM-EELS and other spectroscopic imaging techniques.

Over the past two decades, the ML ecosystem has been growing exponentially, with the new network architectures, methodologies, etc. becoming available almost monthly. As a natural sequence, electron microscopy community started to adopt these methods for applications such as semantic segmentation of images, unsupervised learning over spectral and imaging data, and multiple other applications. Several comprehensive reviews on ML based analysis of STEM – EELS data has become available recently[21,22]. However, the preponderant paradigm in the field is the classical imaging and hyperspectral imaging during the experiment, with subsequent analysis of the resultant data sets. Despite the introduction of new sampling methods such as compressed sensing[23] and several early successes in active learning in STEM[24,25], the amount of work on active learning is so far very limited.

The progress of big data methods in areas such as medical imaging[26], robotics vision[27], and autonomous driving[28] brings forth the question as to whether similar supervised ML methods can be useful in STEM. Here, we pose that currently we are limited by the throughput of the data generation in STEM associated with the fundamental physical limits of the tool. This means that we cannot follow the classical ML paradigm of getting more data sets, but rather must focus on getting data smarter. Given the typical time scales involved in the acquisition of the STEM imaging and spectroscopic data, it requires careful considerations of both the timing and the needed computation and data transfer capabilities.

Here we perform such analysis and report the integration of STEM with remote HPC in academic lab. Given universal availability of cloud-based computational resources and broad availability of STEMs, TEMs, and SEMs, this can be of interest for broad community.

## II. Two faces of big data in STEM

Prior to discussing the instrumental integration and capabilities, we believe that it will be beneficial to better pinpoint the problem we are facing in the STEM experiments, specifically the balance between data and machine learning, By now, it is universally accepted paradigm that success of ML based workflows hinges on the availability of relevant training data[29], with the main progress of the ML community inevitably connected with increasing datasets[30] (from Iris and MNIST to CIFAR and whole Wikipedia) and model sizes.

However, this progression becomes more nuanced when applied to the electron or any other microscopy data[31]. Here, we encounter two main embodiments of big data: one being data aggregated from multiple experiments and the other being large volumes of data from a single experiment. While both can be "big" in terms of volume, they are fundamentally different in terms of contexts in which they are defined and opportunities they open.

Data from multiple experiments as proposed by Ede[32] and Agar[33] are in principle as varied as the chemical space of materials themselves, encompassing different materials and imaging conditions. This diversity means the dimensionality of the generating space is vast. The reason we can train models for tasks like semantic segmentation, despite the vastness of materials' chemical space, is that most materials conform to simple structures of several main classes. Similarly – an atom is an atom - and an image of and atom is ideally a delta function convoluted with probe shape. These two main factors make the model training possible.

At the same time the data from a single experiment, though voluminous, usually explores just a single point within this immense chemical space, often under a single set of instrument settings (and sometimes under controlled sweep of external parameters - like videos of dynamic chemical and temperature induced processes). However, even the chemical spaces of a single sample (meaning conditioned on the composition and synthesis conditions) can be vast – e.g. types of defects clusters, etc. Hence, the big data task of a single imaging experiment that can be either full mapping of the chemical space of the system or structure property relationships via STEM-EELS can be daunting,

Traditionally, the only method to obtain such data was through exhaustive grid measurements—a time-consuming and resource-intensive approach. For expensive spectroscopies, exhaustive grid measurements are not feasible on a large scale, constrained by experimental times and probe damage.

Supporting this statement, our ventures into active learning methods[34–37] for mapping structure-property relationships utilizing pre-acquired datasets, have unveiled that the most interesting regions are often identified within just 20-30 steps out of 10,000 sampled locations. This efficiency can be enhanced with strategic initial point selection. At first glance, this might seem disappointing, suggesting that pre-acquired datasets are superficial. However, this also suggests that active learning strategies deployed on active microscope will therefore allow exploring much bigger regions of chemical space of the system and structure-properties relationships than possible using classical sampling strategies.

## III. ML workflows for STEM

First, we consider the spectrum of active learning tasks emerging in the context of tools capable of imaging and spectroscopy data acquisition. Currently, the predominant approach in the field involves forward workflows, wherein observed imaging data guide the selection of locations for spectroscopic measurements, as implemented in mainstream software such as DigitalMicrograph (DM, Gatan Inc., Pleasanton, California, USA). Inverse workflows, on the other hand, aim to discover microstructural elements that maximize certain aspects of spectral response.

The forward method include:
- **Object finding using simple tools**[38]: This might involve identifying objects or patterns within data using straightforward algorithms, possibly enhanced by hyperparameter optimization to improve performance.
- **Unsupervised discovery and sampling in the feature space**[39]: This method doesn't rely on predefined labels or outcomes but explores the data to find structure or patterns through techniques like clustering or dimensionality reduction.
- **Inference via pre-trained Deep Convolutional Neural Networks (DCNNs)**[39,40]: Using models trained on vast datasets to make predictions or analyze new data.
- **Human in the loop updates**[37,41–43]: Integrating human feedback into the model to refine outputs continuously, enhancing the learning process or model accuracy.

Inverse method include:
- **Deep Kernel Learning (DKL)**[44,45]: Combining Gaussian processes and neural networks to learn complex functions, often used in scenarios where understanding the underlying relationships between structure and property. Similar can be accomplished with RL, bandits, etc. To the best of our knowledge, the latter has not yet been attempted.
- **Reinforcement Learning (RL), Multi-Armed Bandits**[46]**:** These are techniques where systems learn to make decisions by trying different actions and learning from the consequences of each action, effectively learning from trial and error.

**IV. STEM timing and STEM – HPC integration**

Integration of the ML workflows on the active instrument requires analysis of the several key parameters, namely the latencies of data acquisition, transfer to and from the cloud resource, and compute, as well as associated computational requirements. For the latter, we note that modern cloud architectures allow almost unlimited capability to scale the available computational resources. Hence, below we focus on the integration and analysis of transfer rates.

## IV.a. Instrument HPC integration

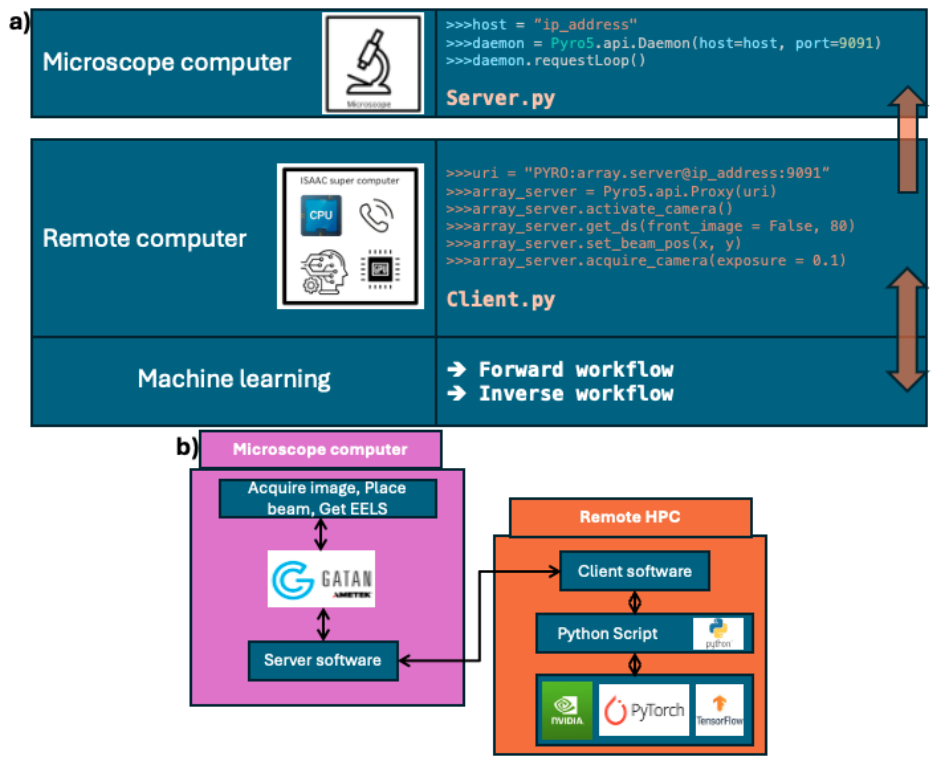

**Figure 1. (a).** Schematic illustrating the server-client-algorithm framework for software layer interaction, including the commands to initiate server and client processes and their communication. **(b).** Diagram depicting the interaction between hardware and software layers, focusing on image acquisition, beam placement, and EELS data collection. The microscope computer uses server software for these tasks, while the remote HPC utilizes client software, Python scripts, and machine learning libraries like PyTorch and TensorFlow. Together, (a) and (b) provide an elaborated view of the software abstraction and its integration with hardware components.

The control of the new TEM/STEM Spectra 300 was achieved through a server software implemented into DigitalMicrograph. Figure 1 a) gives a pictorial overview of the functionalities of the server software which enables the workflows. Note that full analysis here can be complicated since commercial instruments have their internal architecture and compute, and the user access is open only at certain command level. Hence for practical reasons we time the connection and execution time from instrument to HPC and try to define some of the contributing elements. We however do not quantify the instrument-specific execution and assume that it is faster than other limits. 70% of all the STEM in the world have a Gatan DigitalMicrograph software to collect eels. This makes our work directly accessible to 70% of STEM community through DM software.

As illustrated in figure 1 b) server-based application, which establishes direct interaction with the microscope's application programming interface (API). The software has functionalities such as acquisition of scanned image, monitoring of the current position of the beam, placing the beam to predetermined coordinates, retrieval of Electron Energy Loss Spectroscopy (EELS) data, and gathering associated parameters (exposure, resolution, pixel-time etc). The scripting software is open-sourced for users to add/suggest more functionalities.

### IV.b. Timing

At this point EELS spectra can be acquired at any point with sub-pixel accuracy of the overview image. While a grid scan of such an image would (depending on pixel size) require 30 minutes (80x80 pixels) to 4 hours (256x256 pixels), the relevant pixels (interface surfaces, or dislocations) are usually in the order of a 1000 pixels even for a 256x 256 pixel image and thus an acquisition time of 10 min is needed. Therefore, even with the added overhead of 0.2 seconds, a speed-up of a factor of 100 is achieved for this kind of operation.

| Actions | Local computer | Remote computer |
| --- | --- | --- |
| Establish connection | ~ 20 us(microseconds) | ~ 74.4 us(microseconds) |
| Activate camera | ~ 22.9 ms | ~ 39.7 ms |
| Acquire digiscan image 128*128 | ~ 3.91 s | ~ 4.12 s |
| Read beam position | ~ 1.48 ms | ~ 2 ms |
| Set beam position | ~ 6.97 ms | ~ 9.32 ms |
| Acquire eels exposure(1 sec) | ~ 3.8 s | ~ 7.02 s |

### V. Examples of integrated workflows

With these, we deploy and time several prototypical workflows, including simple edge finding, deep convolutional neural networks (DCNN), and deep kernel learning (DKL). These three examples represent the simple image analysis, direct, and inverse workflows and at the same time offer almost unlimited potential for real-world applications.

### V.a. The direct object detection via edge filter

To illustrate the simple image analysis-based workflow, we have chosen the GaAs-Au nanowire sample. Gallium arsenide (GaAs) nanowires are of particular scientific interest because GaAs has an immediate potential for applications in transistor, light-emitting-diode, photodetector, nanolaser, and related technologies[47]. Understanding the growth interface structure between the GaAs nanowire and the catalytic Au nanoparticle is important for ensuring growth of high-quality nanowires[48]. The image contrast is directly related to the square of atomic density at each pixel, giving the common name Z-contrast image. The different structures in the image also poses distinct electronic properties, which can be effectively probed using EELS. An EELS spectrum provides such information as the electronic structure changes that occur at the interface, including changes in the local density of states, band alignment, and charge transfer processes. The interface region may exhibit intermixing or the formation of interfacial compounds, which can be characterized using EELS by analyzing the fine structure of the core-loss edges. Interfaces are often prone to the formation of defects, such as dislocations, vacancies, or impurities, which can significantly impact device performance. EELS can be used to identify and characterize these defects at the atomic scale.

The example of the image is shown in the Figure 2 (a), clearly exhibiting the nanowire and the gold nanoparticle capping one of the edges. A higher magnification image is shown in Figure 3 (b). The EELS measurements at three locations are shown in Figure 2 (c) and shows changes in the plasmonic states of the sample. A low loss spectrum can be interpreted in term of the dielectric function. The spectrum of the GaAs nanowire shows dominantly the volume plasmon peak and a peak due to double inelastic scattering of this plasmon peak. The spectrum from the interface shows an additional peak that is clearly attributed to that interface. Such a peak could also be used as a reward function for active learning.

To demonstrate the remote control of the instrument, we (describe the workflow) As shown in Fig 2 an 128x128 pixel image captured via microscope, which requires approximately 4.5 seconds to acquire from a remote computer. Utilizing edge detection to analyze spectroscopic data across 16,384 pixels, the algorithm identifies interface, effectively reducing the focus to 137 points. This represents a computational reduction to merely 1% of the total area, resulting in a significant efficiency gain where the process now takes 2 minutes instead of the grid-based method taking 80 minutes.

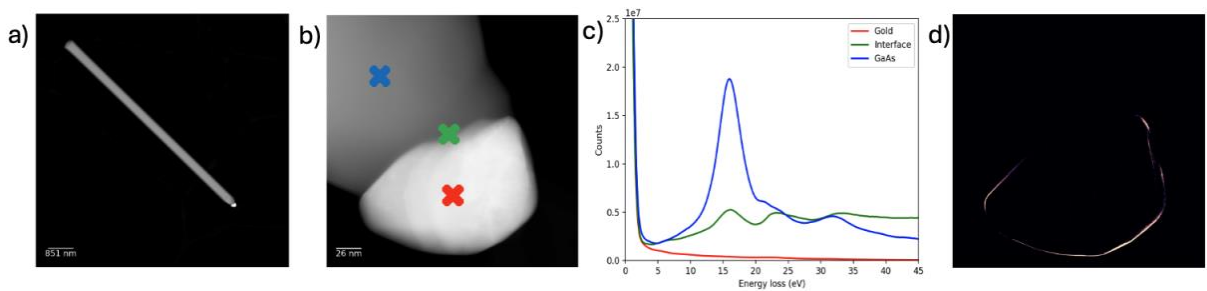

**Figure** 2. **(a).** The High-angle annular dark-field (HAADF) image of the GaAs-Au nanowire sample, **(b).** Higher magnification HAADF exhibiting the interface between gold and gallium-arsenide with three markers, **red**(Au bulk), **blue**(GaAs bulk) and **green**(interface). **(c).** The spectrum collected at the points highlighted in (b). **(d).** Interface region found by edge filter algorithm to conduct EELS.

**V.b. DCNN(ELIT)**

The second broad range of the Ml-based STEM workflow[36,37] is based on the use of the DCNNs for the real-time segmentation of the STEM data, i.e. identification of the atomic positions and identities. Once the atomic positions are identified, a broad number of downstream tasks become possible such as quantitative measurements of interatomic distances, chemical composition mapping, and defect characterization. This process facilitates the detailed study of interfaces and boundaries, critical for understanding and optimizing material properties. Additionally, accurate atom segmentation enhances advanced imaging techniques and supports automated data processing, making high-throughput studies feasible and accelerating material discovery.

These include visualization via unsupervised or supervised analysis[20,49–52], EELS measurements on the chosen atomic groups[37,53–55], electron beam manipulation[56–6364], or direct piping of the data into simulation environments[62]. The fundamental limitation of the DCNN for the real-time analysis of the streaming data from the STEM is the out of distribution shift effects, manifesting as the network performance deteriorating strongly with the changes of effective resolution or sampling compared to the conditions at which training data had been collected. Previously, we have reported the use of the ensemble networks based on collection of U-net architectures as one approach to solve this problem. Graphical processing units (GPU's) are necessary to run the dcnn's, thus our remote software enables the leveraging the compute of external server.

Here, we demonstrate the implementation of the ensemble DCNNs. To show performance on model systems with different lattice symmetries, we use HAADF atomic resolution images of hexagonal graphene (Figure 3a) and cubic strontium titanate (STO) (Figure 3c).

Realtime inference of Ensemble Learning-Iterative Training (ELIT[65]) to the graphene image presented in Figure 3(a) enables the identification of carbon atoms, as depicted in Figure 3(b). When employed on an Nvidia V100 GPU, the initialization of the twenty-model ensemble on the GPU is completed in approximately nine minutes, a process required only once. Subsequently, the ensemble delivers predictions for a 1000x1000 pixel image in under six seconds. One 1000*1000 image acquisition and transfer takes about 5 seconds.

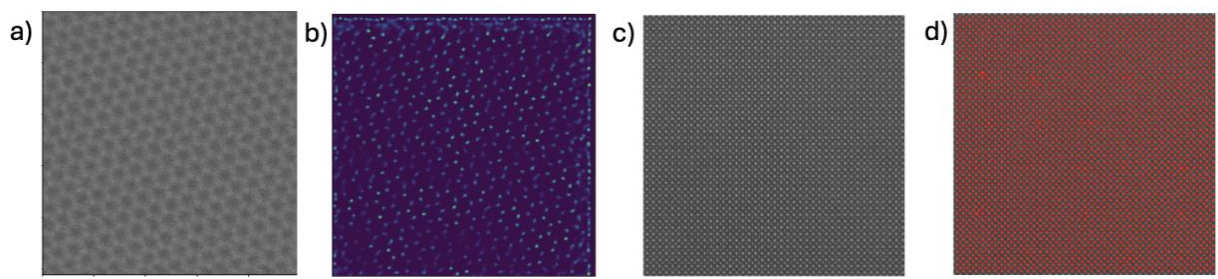

**Figure 3. (a).** HAADF image of graphene, showcasing inherent poison sampling noise **(b).** Detection of atoms on graphene using the ELIT framework, demonstrating the algorithm's capability to accurately identify atomic positions despite the noise. **(c).** HAADF image of strontium titanate (STO) **(d).** Detection of atoms on STO using the ELIT algorithm, illustrating its effectiveness in identifying atomic positions across multiple atomic weights, thus image intensities.

**V.c. DKL**

As the final example of the ML-driven STEM workflow, we reproduced the inverse Deep Kernel Learning based workflow [36,37,41]. In these, the primary task is to sequentially acquire EELS spectra to discover the structural elements in real space that manifest specific spectral signatures of interest. This process involves training a deep kernel learning model on initial patches of data, using a scalarizer to define the measure of interest based on prior knowledge and experimental goals. For this example, we use a mixture of silver nanoplates and gold nanoparticles drop cast onto a lacey carbon film. The volume plasmon peaks of these particles as well as the carbon substrate are at very similar energies. Because gold has the largest response in this energy range (largest relative thickness) we want to define an algorithms that uses the intensity to find those particles.

As described in the work the initial conditions include selecting a scalarizer function and defining the acquisition function that balances exploration and exploitation. Scalarizers can range from simple spectral bandpass filters to complex functions involving peak height ratios or asymmetry. The acquisition functions, such as Expected Improvement (EI) and Upper Confidence Bound(UCB), are dynamically tuned during the experiment to guide the selection of subsequent measurement points. The priors used in the DKL model combine a neural network for feature extraction with a Gaussian Process (GP) for predictive modeling, allowing the DKL model to learn both from structural and spectral data simultaneously.

Figure 4 showcases the automated experimental process facilitated by Deep Kernel Learning[37,41] where (a) is HAADF Z-contrast image of the of silver nanoplates and gold nanoparticles. Figure 4b depicts the learning curve, which provides an insight into model convergence, with (c) showing the selection trajectory of data points during the automated experiment. The training of machine learning (ML) model takes 5 seconds for each iteration whereas eels acquisition takes about 0.5 seconds. 100 steps of exploration took about 10 minutes. Over this time, we can see the model taking measurements which maximize the plasmon peak, sampling in real space on the gold nanoparticles.

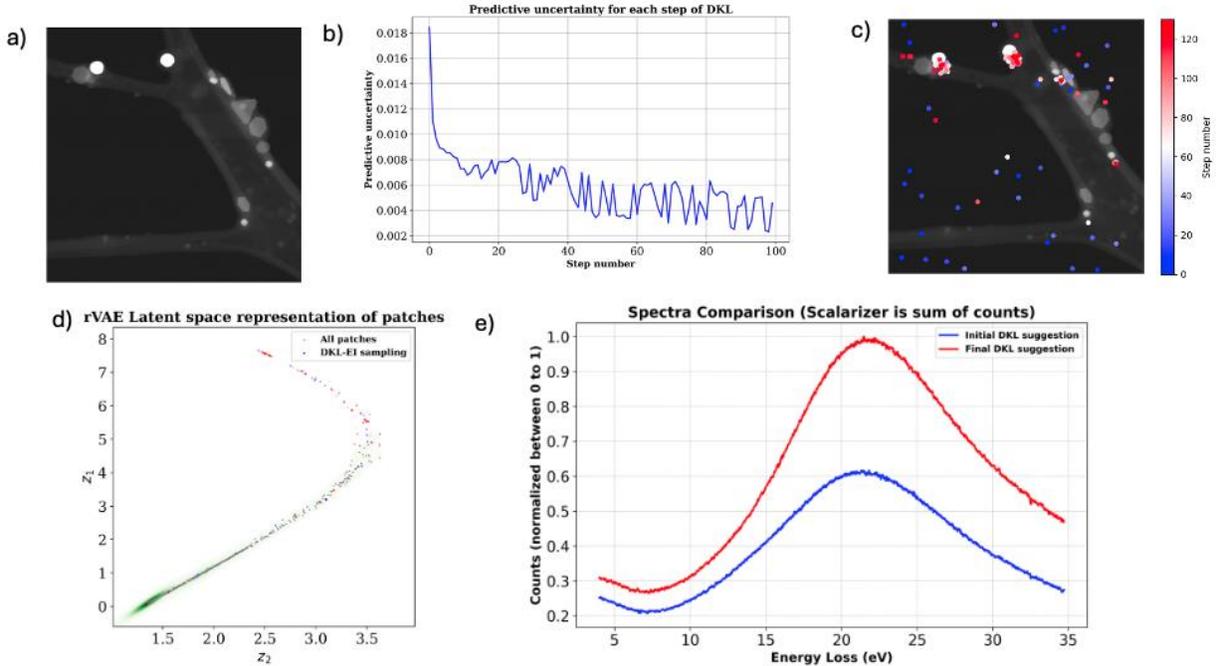

**Figure** 4 **(a).** HAADF Z-contrast image of the of silver nanoplates and gold nanoparticles. **(b).** Predictive uncertainty helps in monitoring the progression of Deep kernel learning. **(c).** Trajectory sampled by dkl in real space. **(d).** Trajectory sampled by DKL in rotational-invariant variational autoencoder(rVAE). **(e).** Comparison of spectra collected during start of experiment and final step of automated experiment.

## VI. Summary and Future work

In this work, we developed a versatile Python server software that facilitates the acquisition of survey images, the precise positioning of the beam, and the collection of EELS spectra. This software's codebase is freely available on GitHub, encouraging the community to build upon and customize the functionality according to their specific needs. To assist users in getting started, we have provided example notebooks within the repository. Despite its capabilities, the software faces a major limitation in data transfer speed, as transmitting a 128x128 image from the microscope computer to the remote server incurs an overhead of approximately 2 seconds. Currently, the server is primarily limited to image acquisition, beam positioning, and EELS data collection. Adding drift correction while acquisition can be easily incorporated using correlation maps of subsequent acquisitions. Future enhancements will focus on incorporating code for EELS alignment, being an open-source project, it invites contributions from the community to expedite these improvements. Furthermore, we speculate that our server software will significantly impact tool capabilities and high-performance computing (HPC) integration. Nearly 70% of STEM microscopes in use today operate with Gatan DigitalMicrograph software, which means our developments have the potential to directly enhance the functionality and efficiency of a substantial portion of the existing microscopy community. While machine learning (ML) algorithms have seen substantial advancements and breakthroughs in various fields, their application in microscopy is hindered by the need for direct control of microscopes from HPC environments. Our server software bridges this gap, enabling more efficient and integrated control, thereby unlocking new potentials for ML-enhanced microscopy.

**Funding**

The implementation of DCNN and DKL workflows on HPC and interfacing between Spectra and HPC (U.P., S.V.K) is supported by AI4Tennessee initiative. The development of STEM control functions (A.H., G.D.) is supported by the U.S. Department of Energy, Office of Science, Basic Energy Sciences, Materials Sciences and Engineering Division.